\documentclass[letterpaper,10pt]{article} 
\usepackage{osameet2}

\usepackage[utf8]{inputenc}
\usepackage{amsmath,amssymb}
\usepackage[font=footnotesize,labelfont=bf]{caption}
\renewcommand{\vec}[1]{\mathbf{#1}}

\begin{document}

\title{Full-Field, Carrier-Less, Polarization-Diversity, Direct Detection Receiver based on Phase Retrieval}%
\author{
    Haoshuo Chen$^{(1)}$, Nicolas K. Fontaine$^{(1)}$, Joan M. Gené$^{(1,2)}$, Roland Ryf$^{(1)}$, \linebreak David T. Neilson$^{(1)}$ and Gregory Raybon$^{(1)}$\linebreak
}

\address{
	$^{(1)}$Nokia Bell Labs, 791 Holmdel Rd., Holmdel, NJ 07733, USA\\
	$^{(2)}$Universitat Politècnica de Catalunya, Jordi Girona 1-3, Barcelona, 08034 Spain	
	}
\vspace*{-0.1cm}
\email{haoshuo.chen@nokia-bell-labs.com}
\vspace*{-0.4cm}
\begin{abstract}
    We realize dual-polarization full-field recovery using intensity only measurements and phase retrieval techniques based on dispersive elements.
   30-Gbaud QPSK waveforms are transmitted over 520-km standard single-mode fiber and equalized from the receiver outputs using 2$\times$2 MIMO.
\end{abstract}
\vspace*{-0.15cm}
\vspace*{-0.5cm}
\section{Introduction}
\vspace*{-0.2cm}
Direct detection communication techniques have obvious limitations compared to coherent detection which can measure the full optical field.
Coherent detection has enabled complex modulation formats, chromatic dispersion compensation, and polarization unscrambling using multiple-input multiple-output (MIMO) processing~\cite{CD}.
However, coherent detection is considered costly and as a result many simplified direct detection receivers have been proposed including the Kramers–Kronig and Stokes space receivers~\cite{KK,Single,Stokes}.

Although these solutions use different receiver/transmitter architectures and various digital signal processing (DSP) algorithms, those that can measure a dual polarization signal and perform MIMO equalization require single-side band heterodyne detection using a continuous wave light carrier, either added at the receiver~\cite{KK}  or before transmission at the transmitter~\cite{Single,Stokes}.
Here, we show a direct detection technique that can measure polarization-multiplexed Nyquist shaped QPSK and QAM signals without any optical carrier at the transmitter or receiver.
It uses phase-retrieval concepts borrowed from image processing and holography~\cite{PRO1} to reconstruct complex-valued signals from two temporal intensity measurements taken before and after a dispersive element~\cite{DISPER}. 
As a demonstration, we measure a polarization-multiplexed 30-Gbaud QPSK signal after 520-km single-mode fiber (SMF) transmission with a line rate of 120-Gbits/s, which has accumulated strong polarization scrambling and chromatic dispersion.
 


\begin{figure*}[!b]
	\vskip -10pt
	\centerline{\includegraphics[width=5.8in]{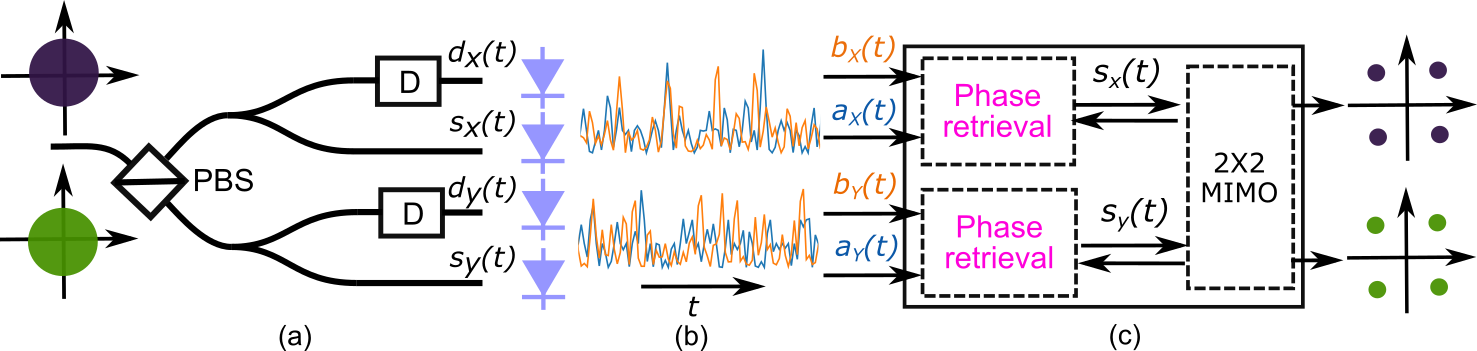}}
	\vspace{-8pt}
	\caption{~(a) Schematic of the carrier-less polarization-diversity phase retrieval receiver for detecting polarization-diversity complex-valued signals based on direct detection (D: dispersive element, PBS: polarization beam splitter),
	(b)~example intensity measured after photodetectors,
	(c)~digital signal processing (DSP) diagram including phase retrieval, equalizer feedback and MIMO processing.
	}
	\label{FIG:1}
\end{figure*}

\begin{figure*}[!t]
	\centerline{\includegraphics[width=6.25in]{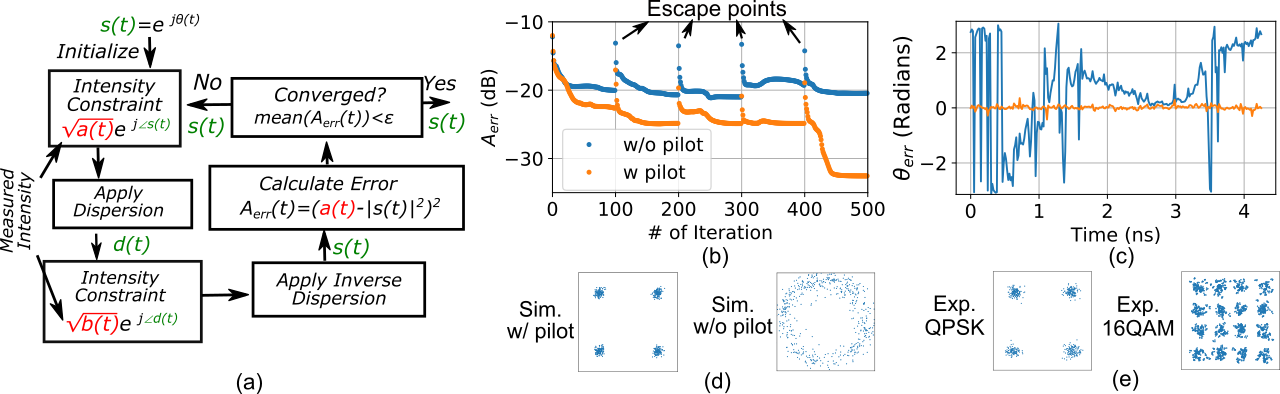}}
	\vspace{-8pt}
	\caption{(a)~Illustration of the Gerchberg–Saxton (GS) algorithm for retrieving complex field, $s(t)$, from two intensity measurements: $a(t)$ and $b(t)$.
		(b)~Phase retrieval error $A_{err}$ versus iteration number w/ and w/o using pilot symbols,
		(c)~phase error over 256 samples after 500 iterations,
		recovered constellations after 500 iterations (d) in simulation and (e) from experimental results using pilot symbols.}
	\vspace{-12pt}
	\vspace*{-0.4cm}
	\label{FIG:2}
\end{figure*}
\vspace*{-0.2cm}
\section{Polarization-Diversity Phase Retrieval Receiver}
\vspace*{-0.2cm}
Fig.~\ref{FIG:1}(a) shows the schematic of the receiver architecture for detecting polarization-diversity complex-valued signals using only intensity measurements.
A polarization beam splitter (PBS) separates the polarization mixed received signal onto the $x$ and $y$ polarizations with respect to the PBS axis.
The intensity of each polarization is measured before, $a_x(t) = |s_x(t)|^2$, and after, $b_x(t)=|d_x(t)|^2$ the dispersive element.
These four real-valued signals, as illustrated in Fig.~\ref{FIG:1}(b) are passed to the DSP (Fig.~\ref{FIG:1}(c)), which is separated into two sections: a phase retrieval front end to recover the full field of each polarization and a 2$\times$2 MIMO equalizer that can compensate for chromatic dispersion, polarization mixing and polarization mode dispersion provided the phase is recovered properly.

The phase retrieval algorithm is applied separately to each polarization and uses a modified Gerchberg–Saxton (GS) algorithm~\cite{PR1,PR2} as illustrated in Fig.~\ref{FIG:2}(a).
The algorithm is  initialized with complex field $s(t)=e^{j\theta_0(t)}$ where $\theta_0(t)$ is a random number.
A improved estimate of the field is constructed from the directly detected intensity and the phase of $s(t)$ (i.e., $\sqrt{a(t)}e^{j \angle s(t)}$ ).
This field is propagated through a dispersive element which is represented by a quadratic spectral phase filter to produce an estimate of the dispersed field, $d(t)$.
Replacing the intensity of $d(t)$ with the second intensity measurement, $b(t)$, improves the estimate of $d(t)$ and it is propagated through inverse dispersion to produce a new guess of the field, $s(t)$.
This process repeats until the mean of the phase retrieval error, $A_{err}(t)= (a(t)-|s(t)|^2)^2$ is minimized below an acceptable error, $\epsilon$. 
The algorithm also includes a spectral constraint to force a rectangular spectrum, and uses feedback from the equalizer such as training symbols/pilot symbols to improve convergence.
In absence of  optical and electrical noise, simulations show QPSK and QAM signals are recovered with only trivial ambiguities such as phase rotation, scaling factor, and linear temporal phase.
With noise, the algorithm can get stuck in many local minima.
Adding 10\% to 20\% pilot symbols and using feedback from the equalizer enables the GS to converge to the correct solution.

\vspace*{-0.2cm}
\section{Simulations and Numerical Procedure}
\vspace*{-0.2cm}



Simulation results for single-polarization 30-Gbaud QPSK signals, are presented in Fig.~\ref{FIG:2}(b-d).
The signals are pulse shaped using a raised-cosine filter with a roll-off of 0.1 and sampled at 60 GSamples/s.
The  OSNR is set to 25~dB and the dispersive element introduces 650 ps/nm chromatic dispersion, which spreads a 30-Gbaud optical pulse over more than 8 symbols.
The entire record of 16384 symbols is processed in blocks of 256 using the overlap save method.

Fig.~\ref{FIG:2}(b) shows the algorithm convergence properties with and without 10\% pilot symbols.
To escape from local minima, the phase of the samples with phase retrieval error above  $\epsilon$ (i.e., $A_{err}(t)>\epsilon$) are multiplied with a random number after every 100 iterations.
With pilot symbols, the mean of $A_{err}(t)$ converges below -30~dB.
Fig.~\ref{FIG:2}(c) shows that the phase error between the recovered and transmitted fields tends to drift slowly without pilots but can be stabilized with pilots.
Likewise, the recovered QPSK constellations with pilots are correct as shown in Fig.~\ref{FIG:2}(d).
Fig.~\ref{FIG:2}(e) shows a back-to-back experiment using the simulation parameters proving that carrier-less QPSK and 16-QAM constellations can be measured using direct-detection.

\vspace*{-0.2cm}
\section{Dual-polarization Transmission Demonstration}
\begin{figure*}[!t]
	\centerline{\includegraphics[width=6.2in]{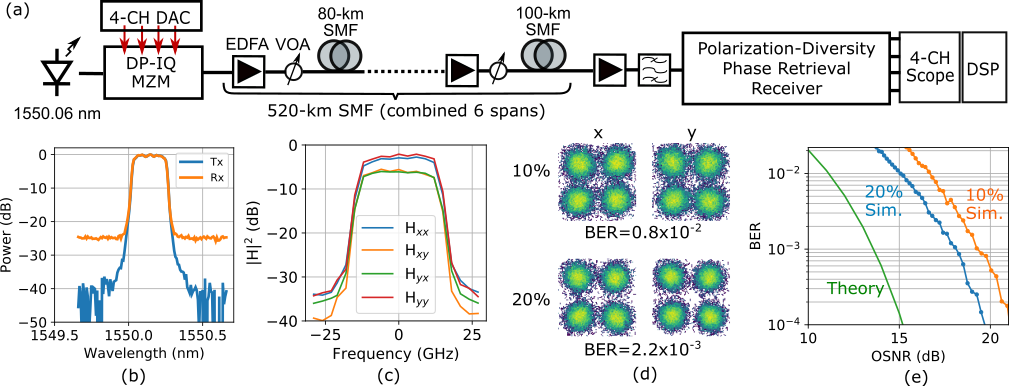}}
	\vspace{-8pt}
	\caption{(a) Experimental setup for detecting polarization-diversity QPSK signals after 520-km SMF transmission,
		(b) measured optical spectra at transmitter and receiver,
		(c) intensity of the estimated fiber channel matrix in the frequency domain,
		(d) recovered QPSK constellations for both polarizations employing 10$\%$ (upper) and 20$\%$ (lower) pilot symbols,
		and (e) BER versus OSNR including theory and simulated results for back-to-back using 10$\%$ and 20$\%$ pilot symbols.}
	\vspace{-10pt}
	\vspace*{-0.4cm}
	\label{FIG:3}
\end{figure*}

\vspace*{-0.2cm}
We transmitted a dual-polarization QPSK signal over a 520-km standard SMF span comprising six SMF spans with a length of either 80 or 100~km to highlight the phase-recovery receiver's ability to handle large amounts of chromatic dispersion and polarization mode dispersion.
Fig.~\ref{FIG:3}(a) shows the transmission experiment.
To generate polarization-diversity 30-Gbaud QPSK waveforms with a spectral roll-off of 0.1 and length of $2^{14}$, an external cavity laser operating at 1550.06~nm was modulated by a dual-polarization IQ Mach-Zehnder modulator (DP-IQ MZM).
It was driven by a 4-channel programmable digital-to-analog converter (DAC).
The input power of each span is adjusted to 0~dBm via a variable optical attenuator (VOA) and the accumulated dispersion is 8921~ps/nm.
After the last span, the optical signal is amplified and spectrally filtered before being detected by the polarization-diversified phase retrieval receiver (Fig.~\ref{FIG:1}(a)).
Two dispersion compensation fibers with a chromatic dispersion around 650 ps/nm are used as the dispersive elements.
After direct detection with four single-ended photo-diodes, signals are sampled by a 4-channel digital sampling oscilloscope at 60 GSamples/s. 

After 520-km SMF transmission, the OSNR at the receiver is 28~dB, and Fig.~\ref{FIG:3}(b) shows the measured optical spectra at transmitter and receiver.
We use the overlap-save method with a 50$\%$ save to enable block-wise phase retrieval with a block length of 1024.
Due to the polarization coupling, the pilot symbols applied for phase retrieval are fed back from the 2$\times$2 MIMO equalizer, as illustrated in Fig.~\ref{FIG:1}(b), and calculated by $\vec{p}=\vec{H} \vec{s}$, where $\vec{p}$, $\vec{H}$ and $\vec{s}$ are the  pilot symbols, estimated fiber channel matrix and transmitted signals, respectively.
Chromatic dispersion from the transmission fiber is compensated prior to the 2$\times$2 MIMO.
Five iterations between phase retrieval and data-aided channel estimation is sufficient to converge the 20 tap equalizer.
Fig.~\ref{FIG:3}(c) shows the intensity of the estimated fiber channel matrix in the frequency domain.
Fig.~\ref{FIG:3}(d) shows the recovered QPSK constellations for both polarizations employing 10$\%$ (upper) and 20$\%$ (lower) pilot symbols with bit-error rates (BER), respectively.
Back-to-back BER versus OSNR curves comparing ideal full-field detection (theoretical) to numerical simulations (without electrical noise/distortions) for the polarization-diversity 30-Gbaud QPSK signals are plotted in Fig.~\ref{FIG:3}(e).
They indicate the OSNR penalty can be $\textless$ 5.5~dB at BER=2$\times 10^{-2}$.
After 520-km (28-dB OSNR), the BER for employing 10\% and 20\% pilot symbols was 0.8$\times 10^{-2}$ and 2.2$\times 10^{-3}$, which is 10-dB worse than the numerical simulation.
Most of this difference is due to the electrical noise from the oscilloscope and lower than optimum photo-currents.

We believe these results can be significantly improved by exploring more advanced phase retrieval algorithms, including  machine learning and other advanced DSP~\cite{PRO1,PRO2}.
Assuming a 20$\%$ overhead for forward error correction (FEC) and 10$\%$ for phase retrieval, a net data rate of 84~Gbits/s is achieved after 520-km transmission.

\vspace*{-0.2cm}
\section{Conclusions}
\vspace*{-0.2cm}
We demonstrated full-field waveform measurement of carrier-less polarization-multiplexed signals using direct-detection/intensity only measurements and phase-retrieval techniques based on dispersive elements.
The direct-detection receiver effectively measures the same thing that a polarization-diversity coherent receiver measures and allows for electronic impairment mitigation using powerful DSP such as 2$\times$2 MIMO processing and chromatic dispersion compensation. 
As a proof-of-concept, we demonstrate polarization-diversity 30-Gbaud QPSK signal detection after 520-km transmission.
Integrated chromatic dispersion can be realized by optical all-pass filters~\cite{SOI1} or Bragg gratings~\cite{SOI2} on chip, permitting future photonic integration of the receiver.

\vspace{-5pt}





\end{document}